\documentclass[a4paper,11pt]{article}
\usepackage{jheppub} 
\usepackage{lmodern}
\usepackage[T1]{fontenc}
\usepackage[latin9]{inputenc}

\usepackage{amssymb}
\usepackage{esint}

\title{Holographic Equipartition and Gravitational Collapse}

\author{Qiao-Jun Cao}

\affiliation{Department of physics, University of Shaoxing, Shaoxing 312000,
China}

\emailAdd{caoqiaojun@gmail.com}

\abstract{
It is argued in the literature that gravity is an emergent phenomenon
and is a statistical tendency for a gravitational system to attain
the maximal entropy state which maintains the holographic principle.
In this paper, we show that gravitational collapse is due to the same
mechanism as the expansion of the universe that the departure of the
surface degrees of freedom and the bulk degrees of freedom in a region
of space drives the evolution of a gravitational system. We also argue
that the ordinary thermal radiation can be taken as a signal for the
existence of some kind of ``confined'' bulk degrees of freedom which
prevent a gravitational system from collapse. When the mount of ``confined''
bulk degrees of freedom is not adequate, the collapse occurs untill
an new equilibrium is attained. Our rusults provide a new paradigm
to understand the gravitational collapse.
}

\begin{document}
\maketitle
\flushbottom

\section{Introduction\label{sec:Introduction}}

There is a strong evidence in the study of spacetime thermodynamics
to suggest that spacetime has internal degrees of freedom, that is
to say gravity and spacetime geometry are emergent phenomena like
fluid mechanics or elasticity\cite{Padmanabhan:2009vy,Padmanabhan:2010xe}.
Since gravity is the most universal force interact between every two
objects, the change in perspective of gravity has important implications
for every dynamic processes in our universe, such as the cosmic expansion
or the stellar evolution.

This perspective is of great significance and has far reaching implications.
In 2010, Erik P. Verlinde proposed an idea that gravity can be explained
as an entropic force\cite{Verlinde:2010hp}. It is really an amazing
idea, it depicts the gravity in a clear physical picture based on
some general assumptions, viz., the spacetime has an microscopic strcture
although we don't know the underlying discrete degrees of freedom,
the gravitational force is caused by changes in the entropy associated
with the locations of material bodies, and inertia is a consequence
of the fact that a material body in rest will stay in rest because
there are no entropy gradients. That is to say, gravity is just a
statistical tendency to return to a maximal entropy state. 

Though some remarkable achievement for the origin of gravity had been
accomplished in Verlinde's entropic interpretation of gravity, some
essential aspects of this frame are still obscure, e.g., (a) The entropic
force $F=T\frac{\Delta S}{\Delta x}$, it requires a non zero temperature
to have a non vanishing entropic force, this means the origin of the
entropic force is closely related to where the temperature comes from.
In Verlinde's paper \cite{Verlinde:2010hp}, the holographic screen
plays an important role in deriving the dynamic equation of gravity
and in discussing the emergence of spacetime. However, there is no
explicit explanation of if we can endow the holographic screen a temperature
and an entropy or not. Horizons have a temperature proportional to
surface gravity which can be identified by the Euclidean continuation
and an entropy measured by its area. It seems thermodynamic quantities
always associate with the horizon, when it comes to the holographic
screen, we definitely need an explanation which will give the entropic
interpretation of gravity a solid foundation? (b) The equipartition
law is a key assumption in deriving Newton's law and Einstein's field
equation. It arises the following questions, what's the implication
of being used to spacetime thermodynamics? can this law applied to
any gravitational system? or what is the context of its application?
(c) In \cite{Verlinde:2010hp}, holographic screen is viewed as the
boundary that separate the spacetime into two parts, on one side there
is spacetime emergent, on the other side nothing yet! It is of no
sense to separate a large room into two parts with a suppositional
screen, and tell someone that one part exists and the other does not.
What is a real edge that separate the emergent spacetime region from
which has not emerged yet? Here, we just point some obscure issues
that should be further investigated, we hope to disscuss it elsewhere. 

Meanwhile, Padmanabhan made a similar discussion about gravity\cite{Padmanabhan:2009kr},
he emphasized that the equipartition law of energy for the horizon
degrees of freedom paly an important role in disscussing the emergence
of gravity, namely the holographic equipartition. Those two works\cite{Verlinde:2010hp,Padmanabhan:2009kr}
are complementary, some obscure aspects about the entropic interpretation
of gravity stated above could be explained by the ideas in \cite{Padmanabhan:2009kr}.
Take the first issue (a) for instance, for a gravitating system enclosed
by a holographic screen, we can construct local Rindler horizons for
every observers placed on the holographic screen who will experience
an acceleration $a$, then, we can attribute an Unruh temperature
$T=\frac{\hbar a}{2\pi k_{B}c}$ to it. That is to say, we can construct
local Rindler horizon for any small patch of the holographic screen
which is associate with the observers who will experience an acceleration
produced by the gravitational body. This explaination fullfills the
logical gap in endow the holographic screen a temperature and an entropy.

Recently, in \cite{Padmanabhan:2012ik}, the accelerated expansion
of the universe is interpreted as the emergence of space and cosmology
driven by the demand of the holographic principle. Specifically, the
holographic principle can be interpreted by the equality of the number
of degrees of freedom in a bulk region of space and the number of
degrees of freedom on the boundary surface, the difference between
the surface degrees of freedom and the bulk degrees of freedom drives
the accelerated expansion of the universe. 

We should also note that there is a hidden hypothesis in \cite{Verlinde:2010hp}
that the maximal entropy state is the state that saturates the holographic
entropy bound\cite{Susskind:1994vu,bekenstein1981universal,Bousso:1999xy,Bousso:1999cb}.
Using the ideas in \cite{Padmanabhan:2009kr} and \cite{Padmanabhan:2012ik},
it is stated that the maximal entropy state maintain the holographic
principle, viz., the number of degrees of freedom in the boundary
is equal to the number of degrees of freedom in the bulk defi{}ned
by the holographic equipartition, that is $N_{bulk}=\frac{\left|E\right|}{(1/2)k_{B}T}=N_{sur}$.
Form this perspective of gravity, we can generalize the observation
in \cite{Padmanabhan:2012ik} and suspect that the dynamical evolution
of every process in our universe related to gravity is governed by
the tendency to saturate the holographic entropy bound. Combining
the ideas of \cite{Verlinde:2010hp} and \cite{Padmanabhan:2009kr},
it is declared that gravity is emergent, it is a statistical tendency
to achieve to the maximal entropy state that maintain the holographic
principle. 

In this paper, we will study the stellar evolution in the spirit of
this paradigm. In section \ref{sec:Expansion-of-Cosmic}, we review
the derivation of Friedmann equation from the novel idea that the
expansion of cosmic space is driven by the demand of the holographic
principle. Our main content is in section \ref{sec:Gravitational-Collapse-and},
we show that the gravitational collapse is due to the same mechanism
as the expansion of universe. It is also argued that the ``confined''
degrees of freedom play an important role in prevent a star form collapse.
Section \ref{sec:Conclusion-and-Discussion} is for conclusion and
discussion. The natural units with $c=1$,$\hbar=1$, and $k_{B}=1$
is used, unless otherwise we reintroduced.

\section{Expansion of Cosmic Space and the Holographic Equipartition\label{sec:Expansion-of-Cosmic}}

Viewed in the perspective of spacetime thermodynamics, the Friedmann
equation of FRW univere is an equation of state, since the Friedmann
equation can be derived from the first law of thermodynamics with
some general assumptions in various gravity theories\cite{jacobson1995thermodynamics,Cai2005,Akbar:2006er},
such as Einstein gravity, Gauss-Bonnet gravity, Lovelock gravity,
scalar-tensor gravity and $f(R)$ gravity, or even in the Ho\v{r}ava-Lifshitz
gravity\cite{Cao:2010xx}. Recently, another approach for reproducing
the Friedmann equation is provided in \cite{Padmanabhan:2012ik},
which extends our understanding of emergence of space and cosmology
dramatically.

Our work is closely related to the ideas in \cite{Padmanabhan:2012ik},
we shall give a brief review on it. The author argued that the difference
between the surface degrees of freedom and the bulk degrees of freedom
in a region of space causes the accelerated expansion of the universe,
the Friedmann equation can be reproduced through a simple equation
\begin{equation}
\frac{dV}{dt}=L_{P}^{2}(N_{sur}-N_{bulk})\:,\label{eq:diff of DOF}
\end{equation}
where 
\begin{equation}
N_{sur}=\frac{4\pi}{L_{P}^{2}H^{2}}\label{eq:sur DOF of U}
\end{equation}
 is the number of degrees of freedom on the spherical surface of Hubble
radius $H^{-1}$that one degree of freedom is attribute to one Planck
area $L_{P}^{2}=\frac{\hbar G}{c^{3}}$, 
\[
N_{bulk}=\frac{\left|E\right|}{\frac{1}{2}k_{B}T}
\]
 is the effective number of degrees of freedom which are in equipartition
at the horizon temperature $T=(H/2\pi)$, and $V=(4\pi/3H^{3})$ is
the Hubble volume in Planck units and $t$ is the cosmic time in Planck
units correspond to the geodesic observers. The effective energy $\left|E\right|$contained
inside the Hubble volume can be taken as $\left|(\rho+3p)\right|V$,
then 
\begin{equation}
N_{bulk}=-\frac{2(\rho+3p)V}{k_{B}T}=-\frac{16\pi^{2}(\rho+3p)}{3H^{4}}\:.\label{eq:bulk DOF of U}
\end{equation}
Substituting Eq.(\ref{eq:sur DOF of U}) and Eq.(\ref{eq:bulk DOF of U})
in Eq.(\ref{eq:diff of DOF}), it simplifies to 
\begin{equation}
\frac{\ddot{a}}{a}=-\frac{4\pi G}{3}(\rho+3p)\:,\label{eq:acceleration equation}
\end{equation}
 which is just the acceleration equation for the dynamical evolution
of the universe. Substitute the continuity equation 
\[
\dot{\rho}+3H(\rho+p)=0
\]
 into the above equation (\ref{eq:acceleration equation}), and integrate
using the de Sitter boundary condition at late times, we finally obtain
\begin{equation}
H^{2}+\frac{k}{a^{2}}=\frac{8\pi G}{3}\rho\:,
\end{equation}
 this is the standard Friedmann equation that governs the evolution
of the universe, the integration constant $k$ can be regarded as
a cosmological constant.

\section{Gravitational Collapse and the Holographic Equipartition\label{sec:Gravitational-Collapse-and}}

The evolution of a star had been extensively studied in the frame
work of general relativity. Generally speaking, a star of a mass under
the Chandrasekhar limit will finally evolve into a stable white dwarf
star, a star weighted between the Chandrasekhar limit and the Tolman-Oppenheimer-Volkoff(TOV)
limit will become a neutron star, and those with masses exceeding
the TOV limit will evolving into a black hole inevitably(see \cite{kippenhahn2012stellar}
for example). Traditionally, gravitational collapse is viewed as a
inward falling process of a star due to its own gravity, form the
point of view of spacetime thermodynamics, it is a thermodynamic process
for a system to attain a equilibrium state. In this section, we will
disscuss the reason of the gravitational collapse from the perspective
of spacetime thermodynamics and it's relation to holographic equipartition.
Here we only study the spherical system for simplicity.

Without loss of generality, for a isotropic spherical body, the spacetime
can be described by the metric 
\begin{equation}
ds^{2}=-e^{2\alpha}dt^{2}+e^{2\beta}dr^{2}+r^{2}d\theta^{2}+r^{2}sin^{2}\theta d\varphi^{2}\:.\label{eq:metric}
\end{equation}
Further, we suppose that the energy-momentum tensor of the matter
in the star has the form of a perfect fluid, viz., $T_{ab}=(\rho+p)U_{a}U_{b}+pg_{ab}$,
where $U^{a}$ denotes the four-velocity of the fluid, and $\rho$,
$p$ are the energy density and pressure, respectively.

Under the framework of general relativity, the structure of a spherically
symmetric star of isotropic material in static gravitational equilibrium
is constrained by the Tolman-Oppenheimer-Volkoff equation 
\begin{equation}
\frac{dp}{dr}=-\frac{(\rho+p)[Gm(r)+4\pi Gr^{3}p]}{r^{2}[1-\frac{2Gm(r)}{r}]}\:,\label{eq:TOV equation}
\end{equation}
where we have defined 

\[
m(r)=4\pi\int_{0}^{r}\rho(x)x^{2}dx\:.
\]
 The condition $m(0)=0$ should be satisfied, which is needed to ensure
that $e^{2\beta}$ is finite. The above equation can also be written
in the form $\frac{dm}{dr}=4\pi r^{2}\rho(r)$, which also satisfies
the relation that 
\begin{equation}
m(r)=\frac{1}{2G}(r-re^{-2\beta})\:.\label{eq:relation of m and beta}
\end{equation}
We will see $m(r)$ is likely to a notion of quasilocal mass that
it's value equals to the Komar mass when the gravitational system
evolute to a final equilibrium state. We should also note that the
proper integrated mass contained in a sphere of radius $r$ is 
\[
M(r)=4\pi\int_{0}^{r}\rho(x)[1-\frac{2m(x)}{x}]^{-\frac{1}{2}}x^{2}dx\:,
\]
which is bigger than $m(r)$, and the difference between these two
quantities can be interpreted as the gravitational binding energy.

Using the ralation of eq.(\ref{eq:relation of m and beta}), the metric
(\ref{eq:metric}) can be expressed as 
\begin{equation}
ds^{2}=-e^{2\alpha}dt^{2}+[1-\frac{2Gm(r)}{r}]^{-1}dr^{2}+r^{2}d\theta^{2}+r^{2}sin^{2}\theta d\varphi^{2}\:.\label{eq:metric B}
\end{equation}

In rest of this section, we shall study the evolution of the isotropic
spherical body under the frame work of spacetime thermodynamics. For
the calculation of it's Unruh temperature, we need to generalize the
definition of surface gravity. Fortunately, the (generalized) surface
gravity definition in \cite{Abreu:2010sc} which has the meaning of
the ``redshifted'' 4-acceleration of the fiducial observer(FIDO)
happen to meet our requirements. This definition has advantages that
can be applied not only to event horizons, but also to FIDOs skimming
along the boundary of a gravitational system. 

Using this proper definition, the surface gravity of the gravitational
system described by the metric (\ref{eq:metric}) can be calculated
as 
\begin{equation}
\kappa=e^{\alpha-\beta}\alpha^{\prime}\:,\label{eq:surface gravity}
\end{equation}
here the $'$ is denote $d/dr$. Then, the correspoinding Unruh temperature
is 
\begin{equation}
T_{U}=\frac{\kappa}{2\pi k_{B}}=\frac{e^{\alpha-\beta}\alpha^{\prime}}{2\pi k_{B}}\:.\label{eq:Unruh temperature}
\end{equation}

The active gravitational mass for a stationary spacetime is called
Komar mass, for the system we discussed above, the Komar mass acquires
the form 
\begin{equation}
M_{K}(r)=r^{2}\kappa=r^{2}e^{\alpha-\beta}\alpha^{\prime}\:.\label{eq:Komar mass of S}
\end{equation}
This could be thought as a generalized form of Newton's law of gravitation
in the circumstance we setted above. 

According to \cite{Padmanabhan:2012ik}, the number of degrees of
freedom on the spherical surface of a gravitational system of radius
$R$ is 
\begin{equation}
N_{sur}=\frac{4\pi R^{2}}{L_{P}^{2}}\:,\label{eq:sur DOF of S}
\end{equation}
and the effective number of degrees of freedom which are in equipartition
at the Unruh temperature $T_{U}$ expressed in eq.(\ref{eq:Unruh temperature})
is 
\begin{equation}
N_{bulk}=\frac{M_{K}(R)}{\frac{1}{2}k_{B}T_{U}}=\frac{4\pi M_{K}(R)}{\kappa}=\frac{4\pi R^{2}}{L_{P}^{2}}\:,\label{DUPLICATE:eq:bulk DOF of S}
\end{equation}
where eq.(\ref{eq:Komar mass of S}) has been used and we reintroduce
the Planch length $L_{P}$ in last equality. We can find that $N_{sur}=N_{bulk}$
is always hold. This result agrees with the statement in \cite{padmanabhan2004entropy,Padmanabhan:2010xh}
that in any static spacetime the equality of $N_{sur}$ and $N_{bulk}$
is maintained.

For the horizonless object we disscussed above, we can argue that
for FIDOs skimming along the boundary of a gravitational system will
see a temperature no less than the local measured Unruh temperature
since the ordinary radiation of the star could be emitted to infinity,
when redshifted to infinity, we get
\[
T\geq T_{U}\:.
\]
Replacing $T_{U}$ by $T$ in eq.(\ref{DUPLICATE:eq:bulk DOF of S})
gives 
\begin{equation}
N_{B}=\frac{M_{K}(R)}{\frac{1}{2}k_{B}T}\leq N_{bulk}\:.\label{eq:B DOF of S}
\end{equation}

I now demonstrate the physical causes that lead to the gravitational
collapse. Assume the isotropic spherical gravitational system evolve
in quasi-static process, namely, the spacetime of the star is not
absolute stationary, but we can still use the metic of eq.(\ref{eq:metric})
to describe it. Using the proposed eq.(\ref{eq:diff of DOF}) to the
isotropic spherical object, viz., insert eq.(\ref{eq:sur DOF of S})
and eq.(\ref{eq:B DOF of S}) into eq.(\ref{eq:diff of DOF}), we
have
\begin{equation}
\frac{dV_{S}}{dt}=-\frac{dV}{dt}=-L_{P}^{2}(N_{sur}-N_{B})\leq0\:,\label{eq:collapse eq}
\end{equation}
note that as the gravitational system collapse the volume of the space
outside it expand, or the change of the volume of the system $\Delta V_{S}=-\Delta V$.
The obove ralation (\ref{eq:collapse eq}) implies that the volume
of the gravitational system $V_{S}$ will decrease as time progresses.
That is: From the perspective of spacetime thermodynamics the gravitational
collapse of a system occurs because it is a non-equilibrium thermodynamic
system, gravitational collapse is a necessary thermodynamic evolution
process for the system to reach equilibrium state when the equilibrium
temperature is just the Unruh temperature without the contribution
of ordinary radiation and the holographic principle attained.

From the point view of quasilocal observer, we can replace $M_{K}(R)$
by $m(R)$ in eq.(\ref{DUPLICATE:eq:bulk DOF of S}), then 
\begin{equation}
\tilde{N}_{bulk}=\frac{m(R)}{\frac{1}{2}k_{B}T_{U}}=\frac{4\pi m(R)}{k_{B}\kappa}=\frac{\frac{1}{2G}(R-Re^{-2\beta})}{\frac{1}{2}k_{B}\frac{e^{\alpha-\beta}\alpha^{\prime}}{2\pi}}\:.\label{eq:bulk DOF of S}
\end{equation}
Assuming the star obeys the holographic principle, namely $N_{sur}=\tilde{N}_{bulk}$,
then we have $\frac{R^{2}}{L_{p}^{2}}=\frac{m(R)}{k_{B}\kappa}$,
which is consistent with the properties of a Schwarzschild black hole
that $\kappa=1/4m(R)$ and event horizon radius $R=2m(R)$. In fact,
If we demand $\frac{r^{2}}{L_{p}^{2}}=\frac{m(r)}{k_{B}\kappa}$ to
hold for any $r$, using the natural units, we get $m(r)=\kappa r^{2}$.
Moreover, by using eq.(\ref{eq:relation of m and beta}), it is easy
to find that $e^{\alpha}=[1-2m(r)/r]^{1/2}$, the metric (\ref{eq:metric B})
becomes 
\[
ds^{2}=-[1-\frac{2Gm(r)}{r}]dt^{2}+[1-\frac{2Gm(r)}{r}]^{-1}dr^{2}+r^{2}d\theta^{2}+r^{2}sin^{2}\theta d\varphi^{2}\:,
\]
which is very similar to the Schwarzschild metric. We can see that
as $T$ tends to $T_{U}$, $m(r)\rightarrow M_{K}(r)$. The above
disscussion indicate that without other interaction except gravity
come into paly, a gravitational system with sufficient mass is tend
to evolve to a maximal entropy state that the Holographic Principle
hold. 

It is also interesting to inquire what mechanism in spacetime thermodynamics
makes a gravitational system stay in stable. Traditionally, a gravitational
system is stable when the gravitational force is counterbalanced by
it's internal pressure, a spherical star in equilibrium is ruled by
the TOV equation (\ref{eq:TOV equation}). The TOV equation (\ref{eq:TOV equation})
is derived by solving the Einstein equation for a general time-invariant,
spherically symmetric metric, and is equivalent to 
\begin{equation}
\frac{d\alpha}{dr}=\frac{m(r)+4\pi r^{3}p}{r[r-2m(r)]}\label{eq:TOV eq before}
\end{equation}
 via the continuity equation $\nabla_{\mu}T^{\mu\nu}=0$. In fact,
eq.(\ref{eq:TOV eq before}) is just the $(rr)$ part of Einstein
equation
\begin{equation}
2r\alpha^{'}+(1-e^{2\beta})=8\pi pr^{2}e^{2\beta}\label{eq:TOV eq as rr part}
\end{equation}
 by using eq.(\ref{eq:relation of m and beta}).

Since the metric of a stable gravitational system is static, in the
spirit of holographic paradigm, it is reasonable to suppose that the
departure of the surface degrees of freedom and the bulk degrees of
freedom divided equally by $\frac{1}{2}k_{B}T$ is balanced by certain
kind of ``confined'' degrees of freedom $N_{c}$ which is invisible
to the observer, that is 
\begin{equation}
N_{sur}-N_{B}=N_{c}\:.\label{eq:confined DOF}
\end{equation}
The obove equation can be rewritten as 
\[
N_{c}=N_{sur}(1-\frac{T_{U}}{T})\:,
\]
 which means the amount of ``confined'' degrees of freedom is a
part of the total degrees of freedom though some kinds of coarse graining
processes and is undetectable when we use the detect ``ruler'' as
$\frac{1}{2}k_{B}T$, it will ``unconfined'' when $T\rightarrow T_{U}$.
This change between the confining and deconfining of degrees of freedom
at $T_{U}$ is probable ralated to a phase transition(for black hole
phase transitions, see \cite{Cao:2010ft} for example and references
therein). It is the confined degrees of freedom that prevent a star
from collapse. When the mount of confined degrees of freedom is not
adequate, the collapse occurs untill an new equilibrium is attained. 

For the spherically symmetric system of isotropic material in static
gravitational equilibrium, it's Komar mass should satisfy some additional
conditions. Insert eq.(\ref{eq:B DOF of S}), eq.(\ref{eq:Komar mass of S})
and eq.(\ref{eq:TOV eq before}) into eq.(\ref{eq:confined DOF}),
we get 
\begin{eqnarray*}
M_{K}(R) & = & \frac{1}{2}(N_{sur}-N_{c})k_{B}T=\frac{e^{\alpha}}{[1-\frac{2m(R)}{r}]^{1/2}}[m(R)+4\pi R^{3}p]=e^{\alpha+\beta}[m(R)+4\pi R^{3}p]\\
 & = & 4\pi e^{\alpha+\beta}\int_{0}^{R}(\rho(x)+3p(R))x^{2}dx
\end{eqnarray*}
which is the expression of Komar mass when the star is in stable.
This expression can also be interpreted as a constraint between $N_{c}$
and $T$, once we have another relation between them from the underlying
theory, we can determine $N_{c}$ and $T$ respectively.

Finally, in the perspective of a quasilocal observer, eq.(\ref{eq:confined DOF})
is replace by 

\begin{equation}
N_{sur}-\tilde{N}_{bulk}=\tilde{N}_{c}\:,\label{eq:other DOF}
\end{equation}
note that $\tilde{N}_{c}$ is different from $N_{c}$. It is easy
to see that the above eq.(\ref{eq:other DOF}) is equivalent to 
\[
M_{K}(R)-m(R)=\frac{1}{2}\tilde{N}_{c}k_{B}T_{U}\:.
\]
 We see that the difference between $M_{K}(R)$ and $m(R)$ is just
the ``confined'' energy. Compare eq.(\ref{eq:other DOF}) with eq.(\ref{eq:TOV eq as rr part}),
using the expression of eq.(\ref{eq:sur DOF of S}) and eq.(\ref{eq:bulk DOF of S}),
we have 
\begin{equation}
\tilde{N}_{c}=\frac{16\pi^{2}r^{3}pe^{2\beta}+2\pi r(e^{-\alpha-\beta}-1)(1-e^{2\beta})}{\alpha^{\prime}}|_{r=R}\:.\label{eq:other DOF result}
\end{equation}
Substitute $\alpha^{'}$ and $e^{2\beta}$ into eq.(\ref{eq:other DOF result})
using eq.(\ref{eq:TOV eq before}) and the relation (\ref{eq:relation of m and beta}),
we finally get 
\[
\tilde{N}_{c}=\left(4\pi r^{2}-\frac{4\pi r^{2}m(r)e^{-(\alpha+\beta)}}{m(r)+4\pi r^{3}p}\right)|_{r=R}\:,
\]
 which agrees with it's definition (\ref{eq:other DOF}) as it should
be.

\section{Conclusion and Discussion\label{sec:Conclusion-and-Discussion}}

In this article, we have studied the gravitational collapse from the
framework of spacetime thermodynamics. We have argued that the appearance
of ordinary thermal radiation makes the observer experience a higher
temperature than it's Unruh temperature, since it is hard for the
observer to distinguish the ordinary thermal radiation from the Unruh
radiation, this effect breaks the holographic equipartition, and there
is a departure from the holographic principle which couses the gravitational
collapse. From this point of view the ordinary thermal radiation is
a signal that the gravitational system is still on the way to it's
final equilibrium state. Gravitational collapse is the thermodynamic
process which makes a system to achieve the equilibrium state when
the isolated matter of the gravitational system is surrounded by a
newly formed horizon and the holographic equipartition recovered. 

We have reviewed that the emergence of cosmic space is also coused
by the difference between the surface degrees of freedom and the bulk
degrees of freedom. This effect can be interpreted that the expansion
of the universe is a response to protect the holographic principle.
Besides the standard evolution of the universe, this approach also
provides a novel paradigm to study other non-standard evolution process
in the universe, such as the inflation\cite{cardenas_inflation_2009}.
It may also help us to understand the origin of the holographic dark
energy\cite{Li:2010cj}.

Our discussion is consistent with the declaration we made in section
\ref{sec:Introduction} that the most stable state for a gravitational
system is the one that maintain the holographic principle. In this
light, the emergence of the holographic dark energy\cite{Li:2004rb}
and the asymptotics of our universe to a de Sitter universe is really
natural.

\section*{Acknowledgments}

I thank Cheng-Zhou Liu and Xuejun Yang for useful disscussions. This
work is supported by the NNSF of China under Grant No.11247306 and
the NSF of Zhejiang Province of China under Grant No.LQ13A050002.

\bibliographystyle{unsrt}
\bibliography{collapse}

\begin{thebibliography}{10}

\bibitem{Padmanabhan:2009vy}
T.~Padmanabhan.
\newblock {Thermodynamical Aspects of Gravity: New insights}.
\newblock {\em Rept.Prog.Phys.}, 73:046901, 2010.

\bibitem{Padmanabhan:2010xe}
T.~Padmanabhan.
\newblock {Lessons from Classical Gravity about the Quantum Structure of
  Spacetime}.
\newblock {\em J.Phys.Conf.Ser.}, 306:012001, 2011.

\bibitem{Verlinde:2010hp}
Erik~P. Verlinde.
\newblock {On the Origin of Gravity and the Laws of Newton}.
\newblock {\em JHEP}, 1104:029, 2011.

\bibitem{Padmanabhan:2009kr}
T.~Padmanabhan.
\newblock {Equipartition of energy in the horizon degrees of freedom and the
  emergence of gravity}.
\newblock {\em Mod.Phys.Lett.}, A25:1129--1136, 2010.

\bibitem{Padmanabhan:2012ik}
T.~Padmanabhan.
\newblock {Emergence and Expansion of Cosmic Space as due to the Quest for
  Holographic Equipartition}.
\newblock 2012.

\bibitem{Susskind:1994vu}
Leonard Susskind.
\newblock {The World as a hologram}.
\newblock {\em J. Math. Phys.}, 36:6377--6396, 1995.

\bibitem{bekenstein1981universal}
J.D. Bekenstein.
\newblock {Universal upper bound on the entropy-to-energy ratio for bounded
  systems}.
\newblock {\em Physical Review D}, 23(2):287--298, 1981.

\bibitem{Bousso:1999xy}
Raphael Bousso.
\newblock {A Covariant Entropy Conjecture}.
\newblock {\em JHEP}, 07:004, 1999.

\bibitem{Bousso:1999cb}
Raphael Bousso.
\newblock {Holography in general space-times}.
\newblock {\em JHEP}, 06:028, 1999.

\bibitem{jacobson1995thermodynamics}
T.~Jacobson.
\newblock {Thermodynamics of spacetime: the Einstein equation of state}.
\newblock {\em Physical Review Letters}, 75(7):1260--1263, 1995.

\bibitem{Cai2005}
Rong-Gen Cai and Sang~Pyo Kim.
\newblock {First law of thermodynamics and Friedmann equations of
  Friedmann-Robertson-Walker universe}.
\newblock {\em JHEP}, 02:050, 2005.

\bibitem{Akbar:2006er}
M.~Akbar and Rong-Gen Cai.
\newblock {Friedmann equations of FRW universe in scalar-tensor gravity, f(R)
  gravity and first law of thermodynamics}.
\newblock {\em Phys. Lett.}, B635:7--10, 2006.

\bibitem{Cao:2010xx}
Qiao-Jun Cao, Yi-Xin Chen, and Kai-Nan Shao.
\newblock {Clausius relation and Friedmann equation in FRW universe model}.
\newblock {\em JCAP}, 1005:030, 2010.

\bibitem{kippenhahn2012stellar}
Rudolf Kippenhahn, Alfred Weigert, and Achim Weiss.
\newblock {\em Stellar structure and evolution}.
\newblock Springer, 2012.

\bibitem{Abreu:2010sc}
Gabriel Abreu and Matt Visser.
\newblock {Tolman mass, generalized surface gravity, and entropy bounds}.
\newblock {\em Phys.Rev.Lett.}, 105:041302, 2010.

\bibitem{padmanabhan2004entropy}
T.~Padmanabhan.
\newblock {Entropy of static spacetimes and microscopic density of states}.
\newblock {\em Classical and Quantum Gravity}, 21:4485, 2004.

\bibitem{Padmanabhan:2010xh}
T.~Padmanabhan.
\newblock {Surface Density of Spacetime Degrees of Freedom from Equipartition
  Law in theories of Gravity}.
\newblock {\em Phys.Rev.}, D81:124040, 2010.

\bibitem{Cao:2010ft}
Qiao-Jun Cao, Yi-Xin Chen, and Kai-Nan Shao.
\newblock {Black hole phase transitions in Ho\v{r}ava-Lifshitz gravity}.
\newblock {\em Phys.Rev.}, D83:064015, 2011.

\bibitem{cardenas_inflation_2009}
Victor~H. Cardenas.
\newblock {Inflation as a response to protect the Holographic Principle}.
\newblock {\em Mod.Phys.Lett.}, A24:2353--2362, 2009.

\bibitem{Li:2010cj}
Miao Li and Yi~Wang.
\newblock {Quantum UV/IR Relations and Holographic Dark Energy from Entropic
  Force}.
\newblock {\em Phys. Lett.}, B687:243--247, 2010.

\bibitem{Li:2004rb}
Miao Li.
\newblock {A Model of holographic dark energy}.
\newblock {\em Phys.Lett.}, B603:1, 2004.

\end{thebibliography}

\end{document}